\documentclass[hyper,letterpaper]{JHEP3} 

\JHEPspecialurl{http://jhep.sissa.it/JOURNAL/JHEP3.tar.gz}

\usepackage{epsfig,multicol,amssymb}

\newcommand\fverb{\setbox\pippobox=\hbox\bgroup\verb}
\newcommand\fverbdo{\egroup\medskip\noindent%
            \fbox{\unhbox\pippobox}\ }
\newcommand\fverbit{\egroup\item[\fbox{\unhbox\pippobox}]}
\newcommand\gtg{\gamma\rightarrow\gamma}
 
\newbox\pippobox

\title{Sensitivity and Insensitivity of Galaxy Cluster Surveys to New Physics}

\author{Joshua Erlich$^a$, Brian Glover$^a$ and Neal Weiner$^b$\\
        $a$ Particle Theory Group, Department of Physics,
College of William and Mary, Williamsburg, VA 23187-8795 \\
        $b$Center for Cosmology and Particle Physics, 
  Dept. of Physics, New York University, \\
New York, NY 10003\\
    E-mail: \email{erlich@physics.wm.edu},
    \email{baglov@wm.edu}, \email{nw32@nyu.edu}}

\preprint{WM-07-107}

\abstract{We study the
implications and limitations of 
galaxy cluster surveys for constraining models of particle physics and
gravity beyond the Standard Model.  Flux limited cluster counts probe the
history of large scale structure formation in the universe, and as such
provide useful constraints on cosmological parameters.
As a result of uncertainties in some aspects of
cluster dynamics, cluster surveys are currently more useful for analyzing 
physics that would affect the formation of structure than physics that
would modify the appearance of clusters.
As an example we consider
the $\Lambda$CDM cosmology and dimming mechanisms, such as photon-axion
mixing.}

\keywords{Large Scale Structure, Galaxy Cluster Surveys}

\begin{document}

\section{Introduction}\label{sec:intro}
In this era of precision cosmology, a wide variety of cosmological
and astrophysical observations are providing strong constraints on
the composition of our universe.  Among these are studies of the
cosmic microwave background (CMB) \cite{WMAP3}, large scale
structure \cite{LSS}, luminosity-redshift curves of Type Ia
supernovae \cite{SNe}, galaxy rotation curves \cite{rot}, and
light element abundances \cite{BBN}.  A relatively consistent
picture of the universe has emerged in which the current universe
is flat $(\Omega=1)$, contains about 20\% of its energy density in
nearly pressureless cold dark matter, about 76\% in
dark energy $(\Omega_\Lambda=.76)$, and the remainder in ordinary
matter described by the Standard Model of Particle
Physics (SM) \cite{WMAP3}. (We take $\Omega_m$ to be the sum of the cold dark 
matter and Standard Model matter, including neutrinos, so $\Omega_m=0.24$
by the above estimates.)
The flatness of the universe and the spectrum of
initial density perturbations is
explained by the paradigm of inflation.  On the other hand,
dark matter and dark energy provide a challenge for particle
physics. The influence of dark matter on galaxy rotation curves, the CMB, 
and most directly in the observed separation of
dark matter and baryonic matter in the ``Bullet cluster'' \cite{bullet},
provides conclusive
evidence that there are new types of particles which have not been
observed in particle physics experiments and are not described by
the SM; and it may be argued that the observation of dark energy in the
expansion history of the universe hints at new gravitational physics.

The incredible precision of
lunar ranging measurements produce some of the strongest
constraints on new gravitational physics \cite{lunar-ranging},
but only on local phenomena that
would be occuring here and now.  The overall expansion history of the universe
constrains the influence of new physics on the largest of scales, 
and indeed the luminosity-redshift
curves of Type Ia supernovae have provided the most direct evidence for
dark energy.  The formation of structure in the universe is also highly
dependent on gravitational and particle interactions,
and since structure has had a relatively long time to form,
galaxy and cluster surveys provide another useful probe of the amount and features of dark matter and dark energy, as well as other new physics.  
The purpose of this paper is to examine the importance of galaxy cluster
surveys in testing of new ideas in
gravitational and particle physics.  (See also \cite{rogerio,jain}.)
It is certainly not a new idea to
use structure formation to constrain cosmological models.  
Indeed, the Press-Schechter formalism for predicting counts of virialized 
objects is more than
30 years old \cite{PS}.  Clusters are the largest virialized objects
in the universe, and as such provide a useful probe of structure formation.
Collisions of hydrogen atoms in the intracluster gas produce X-rays, and
track the gravitational potential well in a cluster \cite{cluster-review}. 
As a result, 
X-ray surveys have provided reliable and complete surveys of galaxy
clusters in various regions of the sky.
Serious studies of the properties of X-ray clusters
for this purpose began in the 1980's
\cite{Mushotzky,David,Markevitch,Arnaud}.  It was suggested by some
groups that cluster
surveys were in conflict with the Concordance Model $(\Omega_m=0.3,\sigma_8
\sim 0.9)$ \cite{Reichart,XMM,Kitzbichler-White,Blanchard-Doupsis}.  
It now seems that the HIFLUGCS cluster survey is in agreement with 
the most recent
Cosmic Microwave Background data ($\Omega_m=0.234$ and $\sigma_8=0.76$)
\cite{WMAP3,Reiprich}.  Our analysis
with the ROSAT 400 Square Degree data set is also in agreement with CMB
data if a large scatter is assumed in the relation between cluster mass
and temperature.  Otherwise, our analysis prefers cosmological parameters
closer to the old Concordance Model.

The particle physics community has not
yet embraced cluster technology for the purpose of testing physics beyond the
standard model.  In large part this is because of limited statistics and 
uncertainties in the theoretical models of structure formation and
cluster dynamics. However, while supernovae are sensitive to the geometry (and opacity) of the universe, while structure growth is sensitive to its clustering properties, these are truly complementary approaches, as argued by Wang et. al \cite{Wang:2007fsa}.
However, since cluster surveys can provide constraints
on new physics complementary to other cosmological constraints, they
deserve to be in the arsenal of the particle physics trade. A purpose of this paper is to review and introduce much of the technology involved to the particle physics community.
As an example of the application of cluster surveys
to particle physics and its limitations, we study the
 significance of current and future surveys for
constraining dimming mechanisms such as the photon-axion oscillation model
of Cs\'aki, Kaloper and Terning \cite{CKT}.  
To motivate consideration of dimming mechanisms, we note that while there
are numerous models of particle physics beyond the SM which 
provide dark matter candidates, the nature of the
dark energy is more of a mystery.  Constraints on the dark energy
equation of state from WMAP and the Supernova Legacy Survey (SNLS 
\cite{SNLS})
suggest that, assuming a flat universe, $w=-0.97\pm0.07$ \cite{WMAP3}, 
where $p=w\rho$ is
the linearized equation of state relating the pressure of the dark energy
fluid $p$ to its energy density $\rho$.  
The value $w=-1$ describes
the vacuum energy, or cosmological constant.  However, naive particle physics
estimates of the vacuum energy are dozens of orders of magnitude too large, 
so it is well motivated to consider alternative models.

If a new pseudoscalar
particle existed with a certain range of mass and axion-type coupling to the
electromagnetic field, then distant objects would appear dimmer than expected
because a fraction of the light emitted by the stars in a galaxy
would have been converted to axions while traversing the intergalactic
magnetic field \cite{CKT}.
It would be necessary to reevaluate
the evidence for acceleration of the universe if the dimming of distant
supernovae could be explained without a cosmic
acceleration.  
At the time when the photon-axion oscillation model was
proposed, a universe without acceleration could not be ruled out if one
allowed for such a dimming mechanism.  
Since that time, new data has provided stronger
constraints on the equation of state parameter of the dark energy,
and a model without acceleration is currently disfavored
\cite{WMAP3}.  However, photon-axion oscillations (or any other viable
dimming mechanism) could still exist, and would lead to
an apparent decrease in the dark energy equation of state parameter, a
possibility which remains open \cite{CKT2}.

A dimming mechanism would also affect flux limited galaxy 
cluster surveys.  Some distant galaxy clusters which would have otherwise been bright enough 
to be detected
in a flux limited telescope, may become too dim to be detected as a result
of photon loss, but it is not a priori obvious what the implications for cosmological analyses would be.  Although the appearance of clusters would be affected by such dimming, such effects can be absorbed into the measured evolution of the luminosity-temperature relation. At any rate, clusters provide an independent test of the nature of dark energy which is complementary to supernovae, and thus potentially constraining of models such as dimming mechanisms.

In the following, we will attempt to provide a thorough review of how cosmology relates to theories of structure formation and our observations. Our analysis
relies on several assumptions regarding cluster luminosities and their 
evolution.
A better 
theoretical understanding of the evolution of cluster properties is desirable
(however, see Ref.~\cite{Maughan}).
On the
other hand, since the apparent evolution of cluster luminosities has been
measured \cite{Vikhlinin,Mullis}, cluster counts provide more direct
constraints on new physics that would affect the formation of
clusters rather than their appearance.
We will use
the 400 Square
Degree ROSAT survey \cite{400d} (hereafter referred to as 400d)
as our primary data set.  
We also use the 400d survey to constrain the standard
$\Lambda$CDM cosmology, which does not require a theoretical understanding
of the mechanisms of luminosity evolution.  

In Section~\ref{sec:structure} we review the  
statistical models of structure formation based on the Press-Schechter
formalism \cite{PS}.  In Section~\ref{sec:dimming} we analyze the possibility
of photon-axion oscillations in light of current galaxy cluster surveys. Interestingly, while supernovae surveys can be dramatically affected by dimming, because the redshift evolution of luminosity and temperature is measured, the studies of cluster count evolution are remarkably insensitive to it (although the total counts, themselves, are).
In Section~\ref{sec:Results} we present our statistical analysis of cluster constraints for standard cosmology, and thus demonstrate the techniques which are simply applied to other theories of modified dark matter or dark energy. 
In Section~\ref{sec:future} we examine the significance of future cluster
surveys for probing new gravitational and particle physics.
We conclude in Section~\ref{sec:conclusions}.

\section{Analytical Models of Structure Formation}\label{sec:structure}
Although the state of the art in structure formation involves elaborate n-body simulations, much can be understood within simple, analytical models.
In this section we summarize the basic theory behind structure formation in
the universe and how the theory is compared with galaxy cluster surveys.
The review is simplified, and does not contain new results, but we hope
it contains enough of the basic ideas so that particle
physicists can easily apply the formalism to constrain new physics.  There
are a number of excellent reviews on structure formation 
in the literature that are substantially more comprehensive than this one,
such as Refs.~\cite{Review}.
Techniques for comparing the models to X-ray cluster data
are somewhat scattered in the literature, though recent cluster surveys
provide useful background with their catalogues, as in 
Refs.~\cite{XMM,400d,REFLEX}.  
Our goal is to simplify the discussion to its bare essentials without
forfeiting too much of the underlying physics, making 
use of the fact that others have performed the complicated simulations
necessary to test both the phenomenological models of hierarchical collapse,
and hydrodynamic scaling relations between
cluster mass and observational quantities like cluster temperature.
Simulations suggest that the simplified models of structure formation
and X-ray cluster dynamics are accurate enough to 
constrain new physics by building the new physics on top of these models.
While simulations are not in perfect agreement with these models
\cite{imperfect},
agreement is good enough that these models serve as a useful tool in studying the evolution of structure.

\subsection{The Press-Schechter Formalism}
The CMB provides strong evidence that the universe was homogenous
to a part in 10$^5$ at the time that atoms formed during
recombination.  However, as the universe
expanded structure formed due to the gravitational collapse of
these small fluctuations into progressively larger objects.  The
precise way in which structure formation occurs is sensitive to
the composition of the universe. Smooth, unclustering dark energy, for example, leads
to a faster expansion of the universe and hinders
formation of structure on large scales.  
Since the evolution of structure 
depends on the composition of the universe, comparison of
models to observations provides an important probe of cosmology.
The Press-Schechter (PS) formalism \cite{PS} provides a simple
model for translating cosmology into number counts for structures
on arbitrary length scales,  as a function of mass and redshift.  Here
we summarize only the main results of this formalism, but there are many
lengthier discussions in the literature
justifying this approach and deriving the relevant results quoted below
({\em e.g.} Refs.~\cite{Review}).

The spherical collapse model of Press and Schechter imagines that
initially overdense regions of the universe collapse with
spherical symmetry,  in an otherwise homogeneous universe.  As the
universe expands, only objects with a density above a critical
value will have collapsed and virialized by any given time. 
For the remainder of this discussion we assume a flat universe with
$\Omega_m+\Omega_\Lambda=1$.
In terms of the cosmological
parameter $\Omega_f(z)$, where \begin{equation}
\Omega_f(z)=(\Omega_m\,(1+z)^3)/(\Omega_m (1+z)^3 +
(1-\Omega_m)) \label{eq:Omegaf}, \end{equation} 
the critical overdensity at the time of
virialization in the linearized spherical collapse model is given
by \cite{K&S}:
\begin{equation}
    \delta_{sc}^v\simeq\frac{3\left(12\pi\right)^{2/3}}{20}\left(1 +
    0.0123 \log_{10}\,\Omega_f\right).
\label{eq:deltascv}
\end{equation}
\noindent 
In the Einstein-de Sitter universe with $\Omega_m=1$, $\delta_{sc}^v\approx
1.69$.  In order to compare with the density field observed today, 
we need to account for the evolution of the
universe.
Since virialized objects are 
nonlinear fluctuations of the background density field, it would seem
difficult {\em a priori} to describe analytically
the evolution of the statistical distibution
of density perturbations.  It was the observation of Press and 
Schechter that since the original spectrum of density perturbations was
approximately Gaussian, and because the precise nonlinear 
evolution of those density
perturbations is unlikely to significantly modify the {\em mass} 
contained in collapsed objects, a linearized approach can be justified
for modeling the distribution of massive collapsed objects.

  In the linear model, density perturbations grow proportional to the growth factor $D(z)$, and thus the linearized overdensity of an object that
virialized at a redshift $z$ has grown by 
 \begin{equation}
\delta_c\left(z\right)=\delta_{sc}^v
\,D(0)D(z)^{-1}, \end{equation} 
where the linear growth factor is defined as \cite{Carroll,Borgani},
\begin{equation}
    D(z)=2.5\,\Omega_m H_0^2 H(z)\int_z^\infty\frac{(1+z')}{H(z')^{3}}dz'\ ,
\end{equation}
\noindent and $H(z)$ is the Hubble parameter at redshift z,
\begin{equation}
H(z)=H_0 \sqrt{\Omega_m(1+z)^3+\Omega_\Lambda}.\label{eq:H}\end{equation}
Below we will use $H_0=100 h$\,km/s/Mpc with $h=0.73$ \cite{WMAP3}.

If one assumes
a Gaussian distribution, the probability of a given collapsed object of mass
$M$ having an  overdensity in the linearized model larger than $\delta_c\left(z\right)$ today is
\begin{equation}
    p(\delta_c\left(z\right),M)=\frac{1}{\sqrt{2\pi}\sigma}\int^{\infty}_{\delta_c\left(z\right)}
    \exp(-\delta^2/2\sigma^2)d\delta.
\end{equation}
By differentiating $p(\delta_c,M)$ with respect to $M$ and dividing by the
volume ($M/\bar{\rho}$) one gets the number density of objects
with mass between $M$ and $M + dM$. The present day mean matter
density of the universe is $\bar{\rho}=2.775\times10^{11}\,
\Omega_m\,h^2\,M_{\odot}\,\rm{Mpc}^{-3}$\cite{RB}.  In this
simplified picture of structure formation, small objects become
bound first, followed by larger structures. 
Galaxy clusters are the largest virialized objects in
the universe in the current epoch, which makes them especially suitable for
study by this approach.

The variance in the distribution of
density fluctuations in the universe,
$\sigma(M)^2=\langle (\delta M/M)^2 \rangle$,  is typically
normalized to spheres of radius $8\, h^{-1}\rm{Mpc}$:
\begin{equation}
    \sigma\left(M\right)^2=\sigma_8^2\, \frac{\int_0^\infty
    k^{2+n}\,T(k)^2\,\left| W\left(k R(M)\right) \right|^2 dk}
    {\int_0^\infty k^{2+n}\,T(k)^2\,\left| W\left(k\,8h^{-1}\,
    \rm{Mpc}\right) \right|^2 dk}
\end{equation}
\noindent where $\sigma_8$ is fit to cosmological
data. The function $W(x)$ is a top-hat filter,\linebreak  
$W(x)=2\left(\sin x - x
\cos x\right)/x^3$, so that $\sigma(M)$ is the variance of the mass 
distribution in spherical volumes of radius
$R\left(M\right)=\left[3M/\left(4\pi\overline{\rho}_0\right)\right]^{1/3}$.
The initial power spectrum is usually assumed to
take the scale-free Harrison-Zel'dovich form $P(k)\propto k^n$, with
$n=1$. As the universe expanded density perturbations grew at
different rates through the radiation dominated era to the present
era. Likewise different size fluctuations crossed the particle
horizon at different times.  The power spectrum therefore varies
from the Harrison-Zel'dovich spectrum in a wavelength dependent
way. Assuming linearity, {\em i.e.} an absence of mixing of modes with 
different wavenumbers, the power spectrum evolved from early times 
may be written as  $P(k) = T(k)^2 k,$\footnote{We absorb the superhorizon 
evolution into $P(k)$; see Ref.~\cite{dodelson} for a discussion.}
where the transfer function, $T(k)$, takes the primordial power
spectrum to the present day. As $k\rightarrow 0$, 
$T(k)\rightarrow 1$ \cite{EH} because
large enough wavelength fluctuations have not crossed the particle horizon
and therefore keep their primordial spectrum. This also means that the
initial time can be taken to be any time before which fluctuations of
interest would have crossed the horizon.  The form of the
transfer function is found by analyzing numerically physical processes 
that would modify the power spectrum over time. For cold dark matter,
Bardeen, {\em et al.} \cite{Bardeen} found,
\begin{eqnarray}
    T\left(k\right)&=&\ln\left(1+2.34q\right)/\left(2.34q\right)
\nonumber \\&\times&
    \left[1+3.89q + (16.1q)^2 + (5.46q)^3 + (6.71q)^4\right]^{-1/4}
\end{eqnarray}
where $q(k)=k/\left(\Omega_m h^2\,\rm{Mpc}^{-1}\right)$ in the absence of
baryonic matter.  To account for baryon density oscillations a ``shape
parameter'' $\Gamma$ is introduced, simply replacing $q(k)$ by 
\cite{Efstathiou-Bond-White92}, \begin{equation}
q(k)=\frac{k}{\Gamma\,h\,\rm{Mpc}^{-1}}, \end{equation}
where \cite{Sugiyama95}, \begin{equation}
\Gamma=\Omega_m h \,\exp\left[-\Omega_b(1+\sqrt{2h}/\Omega_m)\right],
\end{equation}
and $\Omega_b$ is the ratio of the baryon density to critical density.

While the Press-Schechter formalism is remarkably successful in
its comparison to numerical simulations (\cite{PS,SMT,Bond}), 
it has proven to be most powerful as a basis for a
phenomenological approach to modeling galaxy cluster counts.  
One extension to the formalism takes into
account non-sphericity of collapsing objects. Sheth, Mo and Tormen
(SMT) developed a modified PS procedure \cite{SMT} which, allowing for
ellipsoidal collapse, introduces new model parameters which are fit 
to N-body simulations.  In the SMT model,
the mass function is given by

\begin{equation}\label{eq:massfct}
    \frac{dN}{dM}=\sqrt{\frac{2a}{\pi}}\,c\,\frac{\bar{\rho}}{M}
    \frac{d\nu}{dM}\left(1+\frac{1}{(a\nu^2)^p}\right)\exp{\left(-\frac{
a\nu^2}{2}
    \right)},
    \label{eq:dN}
\end{equation}
\noindent where $\nu=\delta_{c}(z)/\sigma(M)$, and the best fit for
the parameters $a$, $c$ and $p$ assuming a standard $\Lambda$CDM cosmology
are $a=0.707$, $c=0.3222$, $p=0.3$ \cite{SMT}.  (By comparison, in the PS
model, $a=1$, $c=0.5$ and $p=0$.)

\subsection{Relating Measured Flux to Cluster Luminosity and Mass}
The Press-Schechter formalism and its extensions 
reviewed above predict the statistical distribution
of massive collapsed objects in the universe as a function of their masses
and redshifts. On the other hand, 
telescopes do not directly measure cluster masses, but rather the flux
and perhaps the
spectrum of light emitted by those clusters in some frequency band
as observed on or near Earth.  In order
to relate the mass function (\ref{eq:massfct}) to observational quantities, it
is necessary to understand the relationships between the mass of a cluster
and observational data.  In this section we describe how properties
of X-ray clusters are related to one another, and how those properties are
then compared with observations.

On average, hydrodynamical models which yield simple scaling relations
between cluster mass and cluster temperature (the $M-T$ relation)
have been proven reasonably successful in comparison with
numerical simulations \cite{RB,Ashfordi}.  
On the other hand, the relation between
the temperature and X-ray luminosity (the $L-T$ relation)
of clusters is sensitive to more
complicated physics such as
cooling mechanisms and the density profile of the intracluster gas,
and is fit by cluster data.
Furthermore, it is now commonly accepted that the $L-T$ relation has evolved
as the composition of radiating cluster gases has evolved \cite{Vikhlinin}.

There are at least two sensible notions of cluster temperature, so it is
important to be precise in terminology.  From here on when we refer to a 
cluster's temperature, $T$, we will mean the temperature of the baryonic gas in
the cluster, as 
is directly measured from the spectrum of light emitted by
the gas in the cluster.  We model the cluster gas as isothermal, which
may not be that good an
approximation for actual clusters \cite{not-isothermal}, although predictions
for number counts are not that sensitive to this assumption \cite{RB}.  
Another notion of cluster temperature is determined by 
the velocity dispersion of the dark matter particles, $\sigma^2=\langle v^2
\rangle$, where the velocity $v$ is measured in the rest frame of the cluster
and the brackets denote the statistical average over dark matter particles.
If typical dark matter particles have a mass $m_D$, then the quantity
$T_D\equiv 
m_D \sigma^2/k_B$ is a measure of the temperature of the dark matter in 
the cluster, where $k_B$ is Boltzmann's constant.  Generally $T_D$ is not
directly related to $T$, 
as the dark matter is not expected to be
in equilibrium with the baryonic matter.  
However, it is often assumed that these temperatures
are similar, or at least proportional to one another, after which a scaling
relation between cluster temperature and cluster mass follows.  

For an isothermal spherical
cluster of dark matter, the density $\rho$ and
velocity dispersion $\sigma$ 
scale with distance from the cluster center $r$ as 
\cite{Binney-Tremaine}: \begin{equation}
\rho(r)=\frac{\sigma^2}{2\pi G_N r^2}, \label{eq:IDF}\end{equation}
with Newton's constant $G_N$.  As mentioned earlier, the assumption of
isothermality may not accurately describe the density profile of the halo, which is a subject of intense study.  A phenomenological density profile which fits better simulations is given by the model of Navarro, Frenk and White (NFW) \cite{NFW}, in which the density profile takes the form,
\begin{equation}
\rho(r)=\frac{\rho_s}{\left(\frac{r}{r_s}\right)\left(1+\frac{r}{r_s}\right)^2},
\end{equation}
where $\rho_s$ and $r_s$ are model parameters.
In the current analysis we assume the isothermal profile, 
Eq.~(\ref{eq:IDF}), for easier comparison to analytic approximations of
scaling relations in the literature.

By considering the evolution of spherical density perturbations, one can 
estimate the density 
of objects which had just virialized at redshift $z$.  The density of
virialized objects may be written in terms of
$\Delta(z)$, the ratio of the 
cluster density to the critical density, $\rho_{crit}=3H^2/8\pi G$.
Assuming unclustering dark energy, and ordinary CDM, a useful
analytic approximation to $\Delta(z)$ was given in Ref.~\cite{BN} for 
flat $\Lambda$CDM cosmologies:
\begin{equation}\Delta\left(z\right)=18
\pi^2+82(\Omega_f(z)-1)-39(\Omega_f(z)-1)^2, \end{equation}  
where $\Omega_f(z)$ is given by Eq.~(\ref{eq:Omegaf}).
A scaling relation between the velocity dispersion and the cluster mass is
obtained by approximating 
the mass of a spherical cluster which virialized at redshift $z$ to
the mass obtained by integrating Eq.~(\ref{eq:IDF}) to a radius such that the 
mean density is given by $\rho_{crit}\,\Delta(z)$, with the result
\cite{BN},
\begin{equation}
\sigma^2\sim M^{2/3} H(z)^{2/3}\Delta(z)^{1/3}. \label{eq:sigma-scaling}
\end{equation}
Here, $H(z)$ is the redshift-dependent Hubble parameter, Eq.~(\ref{eq:H}).
Assuming the baryonic gas in a cluster has temperature proportional to
the dark matter velocity dispersion $\sigma^2$, 
it follows that the cluster
temperature, $T$, scales with cluster mass, $M$, and redshift, $z$, as in
Eq.~(\ref{eq:sigma-scaling}):
\begin{equation}
T\sim M^{2/3} H(z)^{2/3}\Delta(z)^{1/3}. \label{eq:MT1}\end{equation}
In practice, simulations are used to determine 
the constant of proportionality $T_{15}$ defined through \cite{XMM},
\begin{equation}
    T=T_{15}\left(\frac{h}{0.73}\right)^{2/3}\left(\frac{\Omega_m\Delta\left(z,
\Omega_m\right)}{178\, \Omega_f(z)}\right)^{1/3}
    \left(\frac{M}{10^{15}M_\odot}\right)^{2/3}\left(1+z\right)\ ,
    \label{eq:M-T}
\end{equation}
where $\Omega_f(z)$ is given by Eq.~(\ref{eq:Omegaf}), and
$M_\odot$ is the solar mass.  The normalization 
factor 178 is approximately 
the overdensity of a just-virialized object
({\em c.f.} Eq.~(\ref{eq:deltascv}) in the linear model).\footnote{
Refs.~\cite{XMM,Oukbir-Blanchard} 
define $\Delta(z)$
as the contrast density with respect to the {\em background} 
density at redshift $z$.
As in Ref.~\cite{BN}, we 
are defining the contrast density with respect to the {\em critical} 
density, $\rho_{crit}=3H^2/8\pi G$.  
This is the origin of the different scaling relations as written
in Ref.~\cite{BN} and in 
Refs.~\cite{XMM,Oukbir-Blanchard}.  Physically, they are
equivalent, assuming an isothermal density profile.}

Different simulations determine a variety of values for $T_{15}$, which leads
to some ambiguity as to the most accurate normalization 
for the $M-T$ relation.
Typical values are $T_{15}\approx 4.8$ keV and $T_{15}\approx
5.8$ keV \cite{BN}.  We do fits for various values of $T_{15}$ to
gauge the errors associated with the uncertainty in the $M-T$ relation.

With a relation between cluster temperature and cluster mass in hand, the
SMT mass function, Eq.~(\ref{eq:massfct}), can be used to determine
how many clusters of a given temperature are expected per unit
volume
of the sky as a function of redshift.  However, telescopes often have poor
spectroscopic resolution, so that in many X-ray cluster surveys it is difficult
to accurately determine cluster temperatures.  Furthermore, telescopes are 
unable to
observe arbitrarily dim objects, {\em i.e.} they are flux limited.  Hence, in 
order to use the Press-Schechter formalism to predict observed number counts 
of galaxy clusters it is still necessary to relate the cluster temperature to
observed flux.  Such a relationship comes in the form of the elusive $L-T$
relation \cite{Markevitch,Vikhlinin}.  
There are a number of complications in predicting, and in making 
practical use of, the $L-T$ relations which appear in the literature:
({\em i})~Surveys often quote fluxes in some frequency
band, not the bolometric ({\em i.e.} total) flux.  
X-ray telescopes are sensitive to light with frequencies of
fractions of a keV to tens of keV, though
not always in precisely the same frequency band.
({\em ii})~The 
measured frequency band is specified in the telescope's reference
frame, so redshifting of the sources affects the fraction of the total
luminosity observed in the specified frequency band.  ({\em iii})~There is 
some scatter
in the $L-T$ data (from which $L-T$ relations are fitted), 
which is due in part to a
complicated cooling process that takes place in many clusters in
the central region of the cluster gas \cite{Markevitch}.  As a result, 
when possible, some surveys remove the central cooling regions
when inferring X-ray luminosities, and some do not. ({\em iv})~In addition, 
there
is relatively strong evidence that the $L-T$ relation has evolved over time
\cite{Vikhlinin} due to changing cluster environments.  ({\em v})~Furthermore, 
when 
inferring luminosities of distant objects from measured fluxes a particular
cosmology must be assumed, and the assumed cosmology may differ from
one quoted evolving $L-T$ relation to another.  

In this paper we focus on the recent 400d
ROSAT survey \cite{400d}, so we will make use of published $L-T$ data
most
easily compared to the cluster luminosities as presented by the 400d survey. 
In particular, the 400d survey quotes X-ray fluxes in the 0.5-2 keV band
including the central cooling regions.  We begin with the $L-T$
relation determined by Markevitch \cite{Markevitch}
from 35 local ($z<0.1$) clusters.  The fitted
power law $L-T$ relation takes the form \begin{equation}
L_{0.1-2.4}^{local}=A_6\left(\frac{T}{6\,{\rm keV}}\right)^B , 
\label{eq:L-T}\end{equation}
with  $A_6=(1.71\pm0.21)\,10^{44} h^{-2}$ erg s$^{-1}$,
and $B=2.02\pm0.40$, where cooling flows were not removed when inferring
either luminosities or temperatures \cite{Markevitch}.  

To study the redshift dependence of the $L-T$ relation, Vikhlinin, {\em et
al.} \cite{Vikhlinin} measured the X-ray temperature and fluxes of 
22 clusters at redshifts $0.4<z<1.3$ and with temperatures between 2 and
14 keV.  The luminosity $L$ inferred from the
flux $F$ depends on the assumed cosmology via, \begin{equation}
L=4\pi\, d_L(\Omega_m,z)^2\,F\,K(z),\label{eq:LvsF}\end{equation}
so the observed redshift dependence of the $L-T$ relation depends on
the cosmology. The K-correction $K(z)$ will be discussed 
below.  The luminosity distance, $d_L$, is given by,
\begin{equation}
d_L(\Omega_m,z)=(1+z)\int_0^z \frac{c\,dz'}{H(z')}. \end{equation}
The integral is the comoving distance between the source and the telescope.
The extra factor of $(1+z)^2$ in $d_L^2$ accounts for the decreased
energy per photon from redshifting of
the source, and the decrease in frequency between photon arrival times, 
as the universe has expanded.
To correctly
interpret the luminosity evolution in different cosmologies, 
Eq.~(\ref{eq:L-T}) 
should be modified, assuming power law evolution, 
with the reference cosmology factored out:
\begin{equation}
L_{0.1-2.4}=A_6\left(\frac{T}{6\,{\rm keV}}\right)^B \,
\frac{d_L(\Omega_m,z)^2}{d_L(1,z)^2}\,(1+z)^\alpha. 
\label{eq:L-Tevol}\end{equation}

Vikhlinin, {\em et al.} found that assuming a 
$\Omega_m=1$, $\Omega_\Lambda=0$ reference
cosmology, $\alpha=0.6\pm0.3$.  It is 
important to stress that a nonvanshing power $\alpha$ does not in itself
imply an evolution of cluster properties, because $\alpha$ is cosmology
dependent.  However, assuming a more realistic $\Omega_m=0.3$, 
$\Omega_\Lambda=0.7$ reference cosmology leads to a still larger power,
$\alpha_{\Omega_m=0.3}=1.5\pm 0.3$ \cite{Vikhlinin}.  Hence, it
seems difficult to argue that the inferred luminosity evolution is due to
a mistaken assumption about the cosmological expansion rate.  We also note
that other surveys find similar results.  For example, the 
XMM Omega project determined
$\alpha=0.65\pm0.21$ \cite{XMM}.  

We need to be able to convert between frequency bands both
as a result of the redshifting of the spectrum, and in
order to compare measurements of surveys in different frequency bands.  
From Eq.~(\ref{eq:L-Tevol}), we can infer a similar relation for
the luminosity in the 0.5-2 keV band (as appropriate for the 400d survey)
in the cluster rest frame if we know the X-ray spectrum.  
The difficulty is that spikes in the spectrum from
atomic excitations contribute significantly to the flux, so some understanding
of the components of the cluster gas is necessary to accurately convert
luminosity in one frequency band to luminosity in another frequency band.
A popular and accurate model 
is the optically thin plasma model of Mewe, Kaastra, Liedahl, and 
collaborators \cite{Mekal}, the so-called Mekal model.  The Chandra
Interactive Analysis of Observations (CIAO) software package \cite{CIAO}
contains code for
the purpose of converting spectra between bands and between reference frames,
and includes packaged spectral models.
The REFLEX cluster survey has also
tabulated conversion factors between luminosities in various frequency
bands \cite{REFLEX} as a function of temperature, for easy comparison of
cluster data to structure formation models without necessitating installation
of the CIAO software.  

We assume that the luminosity
evolution parametrized by the $(1+z)^\alpha$ dependence in Eq.~(\ref{eq:L-Tevol})
is uniform across the spectrum, so that the same power $\alpha$ will describe
evolution of the 0.1--2.4 keV $L-T$ relation as in the 0.5-2 keV $L-T$
relation.  As a test of this assumption we studied the $z>0.4$ cluster 
data by Vikhlinin {\em et al.} \cite{Vikhlinin}, which includes measurements
of flux in the 0.5-2 keV band and bolometric flux.  We checked that 
Vikhlinin {\em et al.}'s fit of $\alpha\approx0.6$ is valid  both with
their measured bolometric fluxes and fluxes in the  
0.5-2 keV band.  
For our fits we use, \begin{eqnarray}
\alpha&=&0.6 {\rm\ \ \ (with\ an\ } \Omega_m=1\ {\rm reference\ cosmology)} \nonumber \\
A_6&=&1.06 {\rm\ \ \  (for\ the\ 0.5-2\ keV\ band)} \label{eq:params} \\
B&=&2.02 \nonumber \end{eqnarray}
as the parameters in Eq.~(\ref{eq:L-Tevol}).

The final factor required to
compare intrinsic luminosity to measured flux is the K-correction.  
The K-correction
converts from the luminosity in a specified 
frequency band in the rest frame to luminosity in the same frequency band in 
the lab frame, as per Eq.~(\ref{eq:LvsF}).  
The K-correction for a source at redshift $z$ in the
frequency band $(f_1,f_2)$ is given by, \begin{equation}
K(T,z)=\frac{\int_{f_1}^{f_2} df\,P_T(f)}{
\int_{f_1(1+z)}^{f_2(1+z)} df\,P_T(f)}, 
\label{eq:K}\end{equation}
where $P_T(f)$ is the rest frame spectral distribution for an X-ray cluster
with temperature $T$, as a function of frequency $f$.
For example, the measured flux $F$ in the 0.5-2 keV band from a cluster
at redshift $z$ with rest frame luminosity $L_{0.5-2}$ and
temperature $T$ is given by,
\begin{equation}
F = \frac{L_{0.5-2}}{4\pi d_L(\Omega_m,z)^2}\,\frac{\int_{0.5(1+z)}^{2(1+z)} df\,
P_T(f)}{\int_{0.5}^{2} df\,P_T(f)}. \label{eq:flux}\end{equation}
The K-corrections are not strongly temperature dependent except at the low
end of typical cluster X-ray temperatures, 
and a simple power law spectrum, 
\begin{equation}
P_T(f)\sim f^{-n},\end{equation}
with index $n=0.5$, is found to give a reasonable fit in the relevant frequency
bands \cite{Jones}.  A comparison of the K-corrections from
the simple power spectrum and from
more precise plasma spectra can be found in Ref.~\cite{Jones}, or
from the documentation for the Sherpa module of the CIAO software 
\cite{Sherpa}.
Since one of our goals is simplicity in comparison of models of new physics
to cluster data we will assume the simple power spectrum in our fits.  
One should keep in mind, however, that without much more difficulty more
accurate K-corrections can be obtained using available software.  

Finally, in order to predict the number of observed clusters it is necessary
to know the probability of the given telescope
detecting a cluster with a given flux.
The selection
function measures this probability, and depends on the particular survey.  The
selection function is often presented as an effective sky coverage area as
a function of flux, but can easily be converted to a detection 
probability.  The ROSAT 400d survey contained a
geometric survey area of $A_{geo}=446.3$ deg$^2$ \cite{400d}.  
The selection probability
is obtained from their tabulated effective sky coverage, $A_{eff}(f)$ as
a function of flux $f$, via, \begin{equation}
P_{sel}(f)= \frac{A_{eff}(f)}{A_{geo}}. \label{eq:Psel}\end{equation}

We are now prepared to calculate the number of clusters we expect to see in
an area of the sky per unit redshift, for a given telescope flux limit.  
The mass function Eq.~(\ref{eq:massfct}) 
gives the distribution of clusters as a
function of mass and redshift.  We convert mass to temperature using
the $M-T$ relation, Eq.~(\ref{eq:M-T}); and then
temperature to measured flux in the appropriate frequency band using the
$L-T$ relation, Eq.~(\ref{eq:L-T}), K-corrected as in 
Eq.~(\ref{eq:K}), 
with parameters specified by Eq.~(\ref{eq:params}).

\section{ Dimming Mechanisms and Cluster Counts}\label{sec:dimming}
\subsection{CKT photon loss mechanism}
As an example of a dimming mechanism we will study the possibility of
photon loss due to photon-axion oscillations (the CKT model \cite{CKT})
as hypothesized by Cs\'aki, Kaloper and Terning.
We first note that both cluster counts and Type Ia supernova surveys extend
to comparable redshifts $z\gtrsim 1$.  Hence, if
the parameters of the CKT model are chosen so as to affect the interpretation
of the supernova data, as in \cite{CKT}, then the same dimming mechanism will
have an affect on flux limited cluster surveys.

The CKT model assumes an axion-like
interaction between a pseudoscalar field $\phi(x)$ and
the electromagnetic field of the form,
\begin{equation}
    {\cal
    L}_{int}=\frac{1}{M_L\,c^2}\,\phi\,\rm{\textbf{E}}\cdot\rm{\textbf{B}},
\label{eq:Lint} \end{equation}
where the dimensionful scale $M_L$ governs the strength of the interaction.

There is significant evidence for an intergalactic magnetic field (IGMF), 
although little is
known about its uniformity in magnitude and direction \cite{IGMF}.
The typical magnitude of the IGMF is estimated to be
$10^{-9}$ Gauss, and it is typically
assumed that the field is coherent to about $L_{dom}\sim 1$ Mpc.
In a background magnetic field the photon-axion
interaction, Eq.~(\ref{eq:Lint}), gives rise to a mixing between
the axion $\phi$ and the electric field.  This mixing causes oscillations
just as mixing between neutrino flavors leads to neutrino oscillations.
In the case of photon-axion oscillations there is one polarization of the
axion and two of the photon, so after traversing enough regions of randomly
oriented magnetic fields a beam of photons will become distributed equally
among the three polarizations.  Asymptotically, the intensity of light
received from a distant astrophysical object will be decreased by
a factor of $1/3$ (in the limit of infinite horizon size).  As
discussed in \cite{CKT}, this effect is approximately described by the
following expression for the probability of a photon to remain a photon
after traversing a comoving distance $r(z)$:
\begin{equation}
    P_{\gamma\rightarrow\gamma}(z)\simeq\frac{2}{3}+\frac{1}{3}\,e^{-r(z)/L_{dec}},
   \label{eq:photonloss}
\end{equation}
where,
\begin{equation}
    L_{dec}=\frac{8}{3}\frac{\hbar^2\,c^6\,M_L^2}{L_{dom}\left|
    \rm{\textbf{B}}\right|^2}.
\end{equation}
The existence of a dimming mechanism like photon-axion oscillations would
modify the interpretation of cluster counts. 
In particular, there would be a reduction 
in the number of distant visible clusters in a flux limited observation.
Dimming can also mimic luminosity evolution, although photon-axion oscillations
cannot explain the observed luminosity evolution.  The fact that $\alpha>0$
in Eq.~(\ref{eq:L-Tevol}) indicates that distant clusters appear more luminous
than nearby clusters, while photon-axion oscillations would have led
to the opposite conclusion.  If we had a theoretically predicted
$L-T$ relation, cluster dimming
would be 
accounted for by including a factor of $P_{\gtg}$ in the $L-T$ relation,
which would become: \begin{equation}
L_{bol}=A_6\left(\frac{T}{6\,{\rm keV}}\right)^B \,
\frac{d_L(\Omega_m,z)^2}{d_L(1,z)^2}\,(1+z)^\alpha
P_{\gtg}
    \label{eq:L-Tdim}.
\end{equation}
Similarly, the measured flux from a cluster in the 0.5-2 keV band,
which was given by Eq.~(\ref{eq:flux}), would become,
\begin{equation}
F = \frac{L_{0.5-2}\,P_{\gamma\rightarrow\gamma}(z)}{
4\pi d_L(\Omega_m,z)^2}\,\frac{\int_{0.5(1+z)}^{2(1+z)} df\,
P_T(f)}{\int_{0.5}^{2} df\,P_T(f)}. \end{equation}
However, since dimming mechanisms would mimic luminosity evolution, it is
redundant to include the factor $P_{\gamma\rightarrow\gamma}$ in 
Eq.~(\ref{eq:L-Tdim}) if the evolution specified by the parameter
$\alpha$ is fit to observations. It is a remarkable fact that, although dimming does affect the appearance of clusters in these surveys, all these effects are absorbed into the $z$-dependence of the $L-T$ relation. Thus, having measured this evolution, these surveys should determine the nature of dark energy independently of dimming.
On the other hand, physics
which would affect the formation of structure rather than the 
appearance of clusters, can be constrained with cluster counts without
a theoretical understanding of the evolving $L-T$ relation.  

Let us note here two important upshots of this fact: first, given a measured 
$L-T$ evolution, implications of cluster counts are independent of dimming 
while supernovae clearly are not. Thus, as data sets expand, comparing these 
two will constrain any anomalous dimming of supernovae. Secondly, distant 
clusters tend to be brighter than they would have been
in the absense of evolution.  With a larger statistical 
sample of x-ray clusters and a better theoretical understanding of $L-T$ 
evolution, this alone may be the strongest constraint on dimming mechanisms.

\section{Results}
\label{sec:Results}
In this section we compare number counts found using the above
model to the 400d ROSAT survey \cite{400d}. The 400d survey
identified 242 optically verified X-Ray sources in the main survey. The
search was done with a flux limit of 1.4 $\times$ $10^{-13}$ erg
s$^{-1}$ cm$^{-2}$ and with a geometric sky coverage of 446.3
square degrees. In order to compare our theoretical number counts
to ROSAT's data, we integrated redshift over a bin size of $\Delta
z=0.1$. Since the 400d survey reported the error bars in their flux 
measurements, we estimated the
error bars on galaxy cluster number counts by counting how many
objects in a given redshift bin would lie below the flux limit when the
measured flux is shifted downward by one standard deviation.  
In the cases
where an X-ray source lied on the boundary of a redshift bin, it
was counted in both of the adjacent redshift bins.  
To account for the scatter in the $M-T$ and $L-T$ relations, 
we included a log normal distribution
in the effective $L-M$ relation ($L(M,z)$): 
\begin{equation}
    P_L(\ln L',z)=
\frac{1}{\sqrt{2 \pi
    \sigma_{\ln L}^2}}\exp\left[\frac{-\left(\ln(L')-\ln(L(M,z))\right)^2}{2
    \sigma_{\ln L}^2}\right] ,
\end{equation}
The effective selection probability $\widetilde{P}_{sel}(M,z)$ of objects
of mass $M$ at redshift $z$ is
a convolution of the survey selection function $P_{sel}(f)$ with the 
distribution in luminosities inferred from the $L-M$ relation $P_L(L,z)$: 
\begin{equation}
\widetilde{P}_{sel}(M,z)=\int_{\ln L_x(z)}^\infty P_L(L',z) P_{sel}\left(\frac{e^{\ln
L'}}{4\pi d_L(z)^2}\right)\,d\ln L', \label{eq:Pselt} \end{equation}
where \begin{equation}
L_x(z)=4\pi d_L(z)^2 f_x \end{equation}
 is the lower limit on the
luminosity at redshift $z$, corresponding to 
the flux limit $f_x$ of the survey, and the argument of $P_{sel}$ is the 
flux expressed in terms
of the luminosity and luminosity distance. We compared
the best fit values of various observables as the assumed 
scatter, $\sigma_{\ln L}$, varied from $0.3$ to $0.7$ \cite{RB,Nord}.  
The results are described below, and can also be
seen in Fig.~\ref{fig:noaxplot} and Fig.~\ref{fig:axplot}.

The number of observed
virialized objects in a redshift bin $\Delta z=0.1$ is then
given by integrating the mass function over objects, weighted by 
$\widetilde{P}_{sel}(M,z)$:
\begin{equation}\label{eq:N}
    N\left(>f_x,z,\Delta z\right) = A_{geo}\int_{z-\Delta z}^{z+\Delta
    z}\int_{0}^{+\infty}dz\,dM\,
r(z)^2 \,\frac{dr}{dz}\,\frac{dN}{dM}\,\widetilde{P}_{sel}(M,z)
,
\end{equation}
where $A_{geo}$ is the geometric
sky coverage in steradians and $f_x$ is the flux limit of
the survey, which feeds into $\widetilde{P}_{sel}(M,z)$ as described above.  

The comoving volume element per steradian is, \begin{equation}
dV(z)=r(z)^2 \left(\frac{dr(z)}{dz}\right)\,dz, \end{equation}
with \begin{equation}
r(z)=c\int_{0}^{z}\frac{dz'}{H(z')}, \end{equation}
and the Hubble parameter $H(z)$ is given by Eq.~(\ref{eq:H}).

It is important to ensure that the lightest mass virialized
object included in $N$ given the assumed $M-T$ and $L-T$ relations 
is cluster sized and not smaller.  Otherwise $N$ contains smaller
objects which are not included in the survey.  This constrains the smallest
$z$ for which this formalism is valid.
In our fits we only include clusters with redshift $z\geq 2$.
The lightest mass virialized object that could have been observed by the 
400d survey, with X-ray
flux limit 1.4$\times 10^{-13}$ erg/s/cm$^2$, assuming $\Omega_m=0.3$ and
$h=0.73$, is around $M(0.2)=1.6\times10^{14} M_\odot$,
which is cluster size.

\subsection{Systematic Errors}
The only errors included in our fits are an estimate of the uncertainty in
low flux cluster counts.  There are in addition
a number of theoretical uncertainties,
some of which we studied by repeating our fits with different parameter
choices.  
Larger normalizations of $T_{15}$ in Eq.~(\ref{eq:M-T}) would
lead to a prediction of brighter clusters, and hence larger cluster
counts.  For a fixed data set, larger $T_{15}$ would then translate into a 
measurement of less structure, corresponding for example to smaller
$\Omega_m$ and/or $\sigma_8$.  Similarly, a larger normalization for
$A_6$ in Eq.~(\ref{eq:L-T}) would imply brighter clusters, with similar
consequences to increasing $T_{15}$.  To examine the uncertainty in 
predictions for cosmological and dimming parameters, we repeated our fits
for typical determinations of $T_{15}$ from simulations, differing by
as much as 20\%.  The uncertainty in $A_6$ is effectively equivalent to
an additional
uncertainty in $T_{15}$ of around 5\%.  The power laws used in the
$L-T$ and $M-T$ relation also have associated errors, and it would be
useful to perform a more complete error analysis.  
Another source of error is 
our assumed power law spectrum used to calculate K-corrections, which
is a worse approximation for low-temperature clusters 
than high-temperature clusters, but we expect that this is not a significant
source of error.
We have also checked that
alternative redshift binning of the data does not significantly change our 
results.  To examine the effect of scatter on our analysis, we 
reduced our assumed scatter from $\sigma_{\ln L}=0.7$ to $0.3$ and found
that the predicted
number counts were reduced by nearly a factor of two, as can be seen below.
This demonstrates the importance of correctly accounting for such statistical
effects.

\subsection{Flux Limited Cluster Counts for Standard Cosmology}
\label{sec:nonewphysics}
\begin{figure}[h]
\centering
\begin{tabular}{cc}
\epsfig{figure=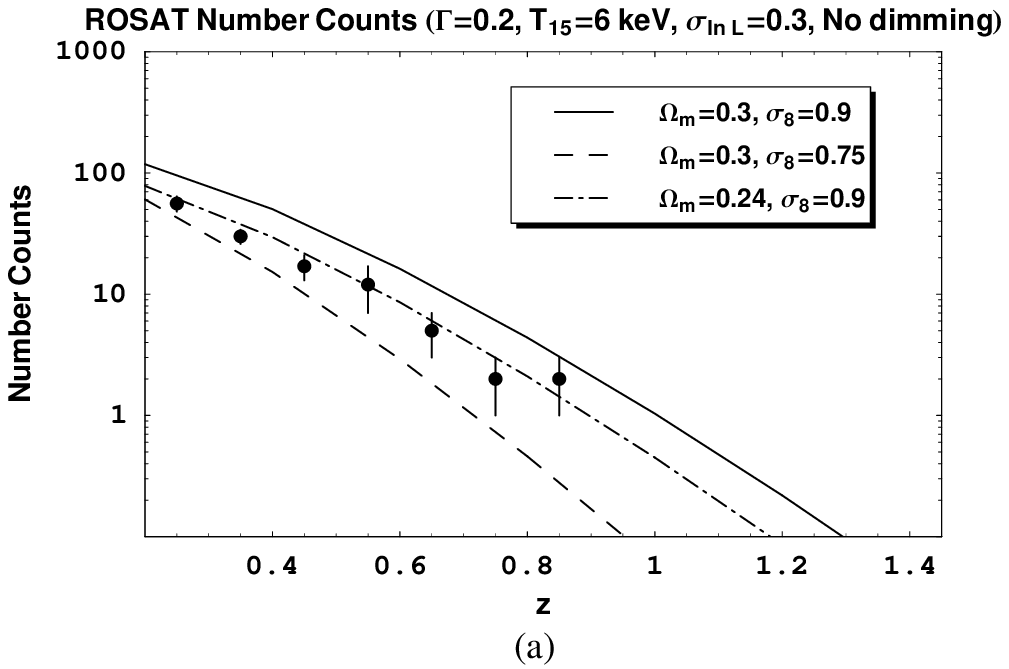,width=3.0in} &
\epsfig{figure=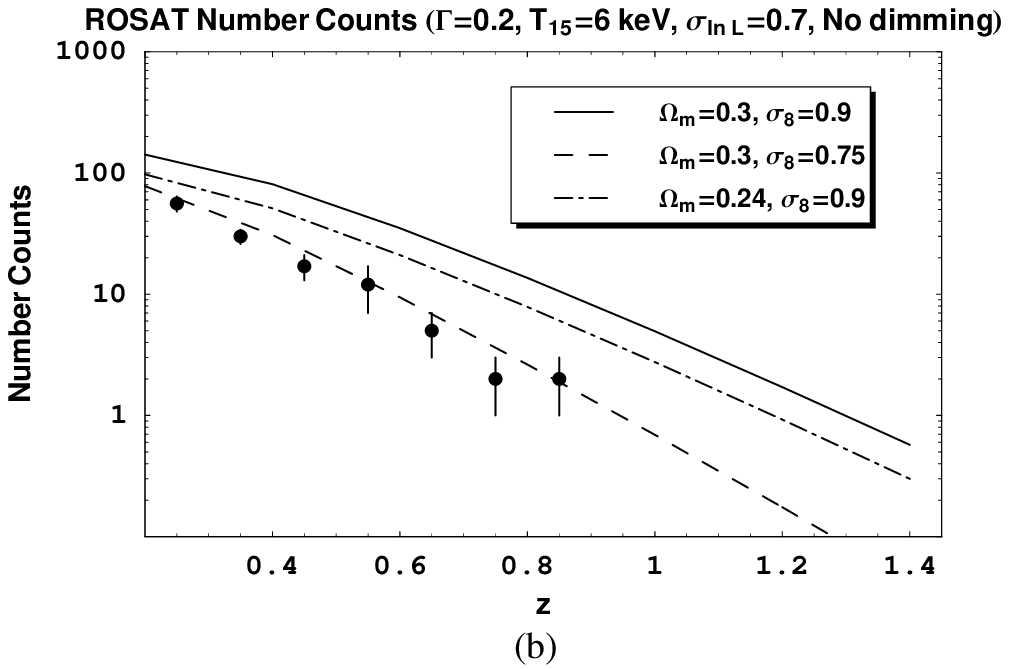,width=3.0in} 
\end{tabular}
\caption{Number Counts
versus redshift for different matter densities ($\Omega_m$) and
matter density fluctuation amplitudes ($\sigma_8$), without photon-axion 
oscillations, with different levels of scatter in the L-M
relation.  
The theoretical predictions correspond to $\Gamma=0.2$ and T$_{15}=6$ keV.
(a) $\sigma_{\ln L}=0.3$.  (b) $\sigma_{\ln L}=0.7$.}
\label{fig:noaxplot}
\end{figure}

Figure \ref{fig:noaxplot} shows our computed number counts and
the 400d survey's observed number counts versus redshift. Three curves
were drawn for different values of $\Omega_m$
$\sigma_8$, with
$\Gamma=0.2$ and T$_{15}=6$ keV.

Larger normalizations for the $L-T$ relation ($T_{15}$) lead to smaller 
predicted values of $\Omega_m$ and $\sigma_8$.
Reducing the assumed level of scatter in the effective $L-M$
relation leads to a decrease in the
predicted number of dim objects observed.  For a given set of observations,
reducing the assumed scatter would then lead to a larger inferred amount
of structure, {\em i.e.} larger  
$\sigma_8$.  For $T_{15}=6$ keV and $\Gamma=0.2$, the
best fit shifts from $\Omega_m=0.209$ and $\sigma_8=0.923$ with 
$\sigma_{\ln L}=0.3$ to $\Omega_m=0.286$ and $\sigma_8=0.731$ 
with $\sigma_{\ln L}=0.7$.

\begin{figure}[h]
\centering
\begin{tabular}{cc}
\epsfig{figure=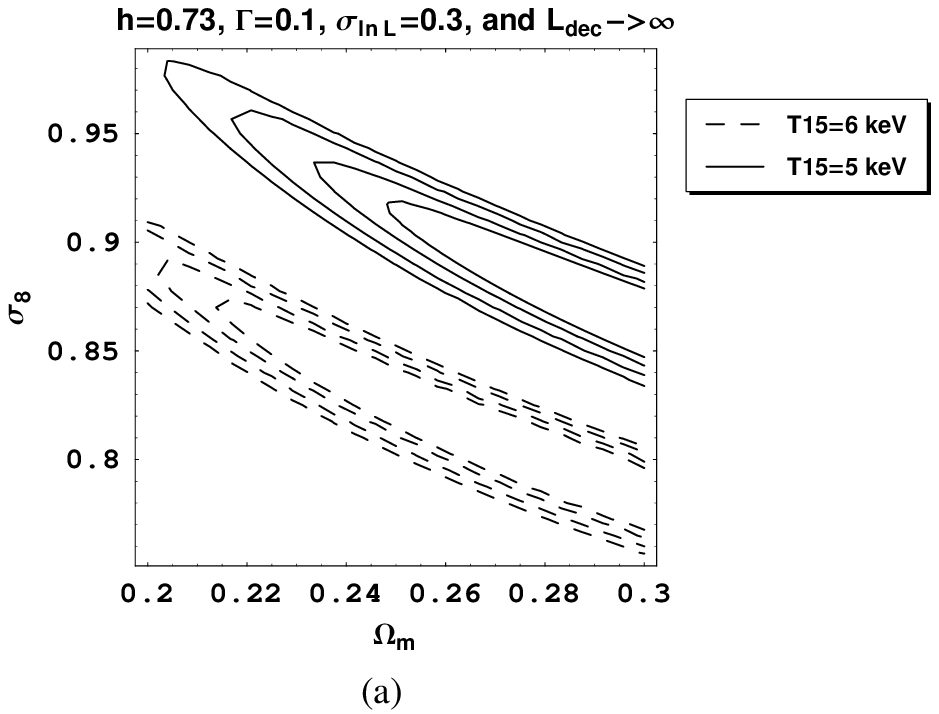,width=3.0in} &
\epsfig{figure=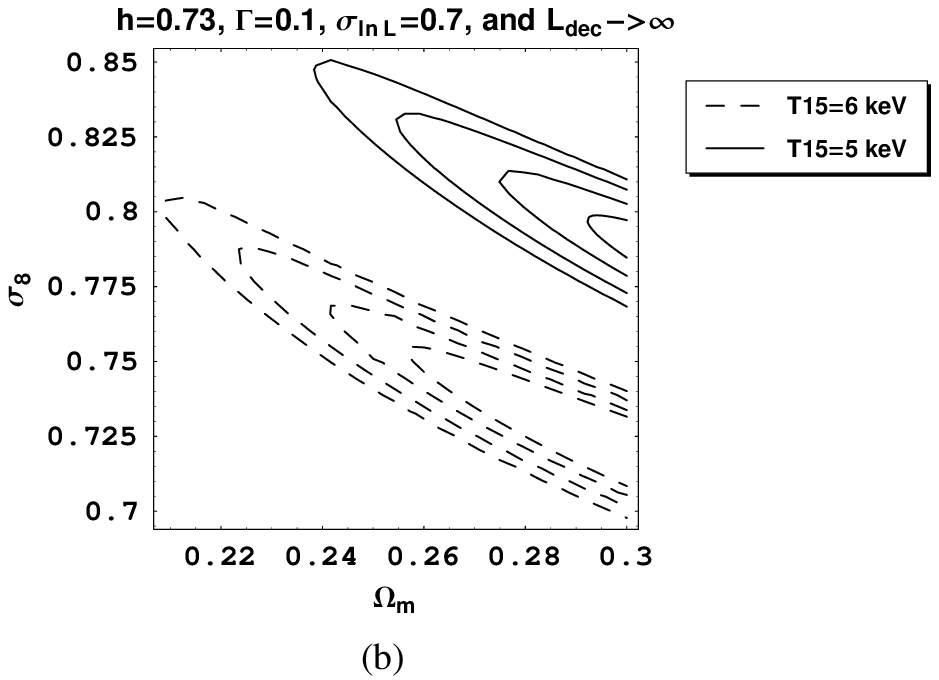,width=3.0in} \\ \\
\epsfig{figure=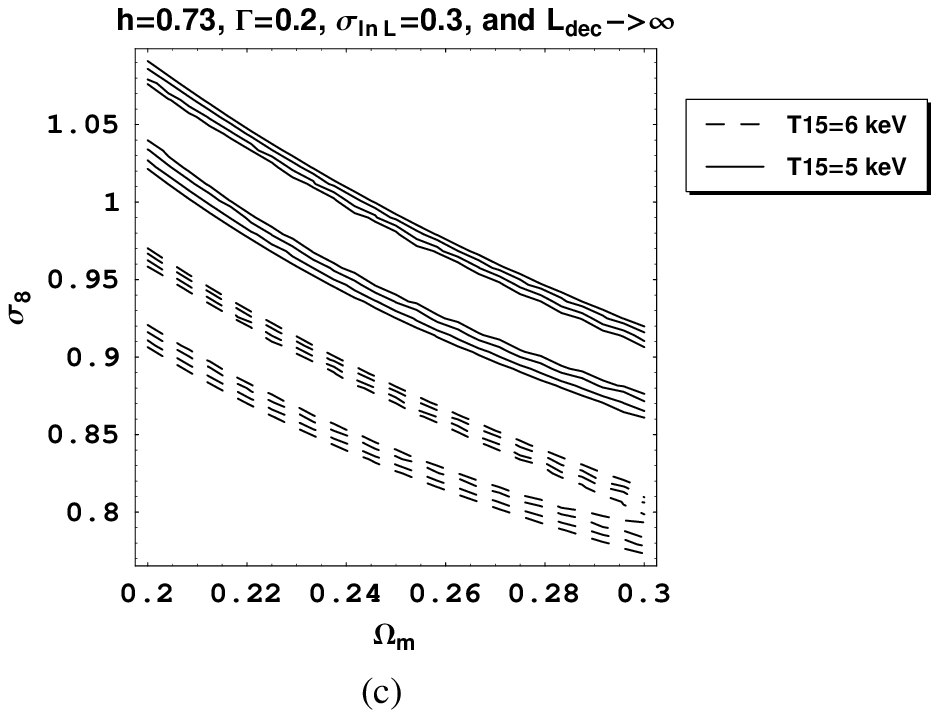,width=3.0in} &
\epsfig{figure=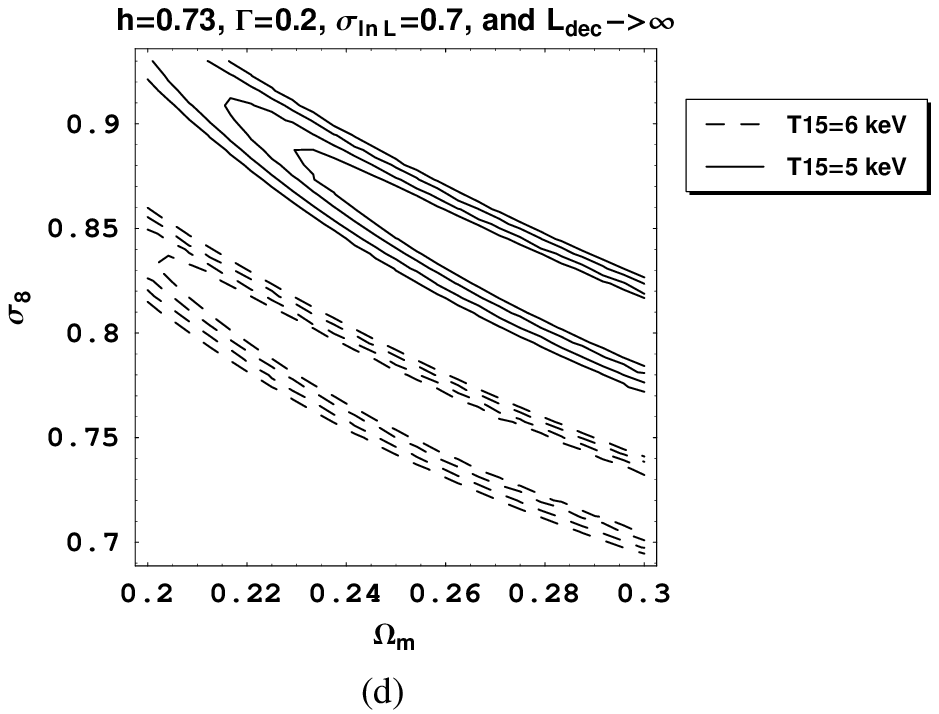,width=3.0in} 
\end{tabular}
\caption{Confidence plot
of $\Omega_m$ and $\sigma_8$ for various choices of model parameters.
The lines represent 68$\%$, 80$\%$, 90$\%$, and
95$\%$ confidence regions.  
(a) $\Gamma=0.1$, $\sigma_{\ln L}=0.3$.  
(b) $\Gamma=0.1$, $\sigma_{\ln L}=0.7$. 
(c) $\Gamma=0.2$, $\sigma_{\ln L}=0.3$.
(d) $\Gamma=0.2$, $\sigma_{\ln L}=0.7$}
\label{fig:cp}
\end{figure}

Figure \ref{fig:cp} shows our
$\chi^2$ analysis for different values of $T_{15}$, $\Gamma$ and 
$\sigma_{\ln L}$.
We only include our estimated uncertainties in the 400d survey number counts
in the statistics.
Notice that for values of $\Gamma$ between 0.1 and 0.2 and
$T_{15}$ between 5 and 6 keV, there is tension between our result and the
best fit WMAP 3-year measurement.  We are consistent with WMAP bounds
if we assume 
large $T_{15}$, small $\Gamma$ and large $\sigma_{\ln L}$.  
Our results are similar to earlier studies
\cite{Reichart,XMM,Kitzbichler-White}, although Reiprich 
\cite{Reiprich} has found that
the HIFLUGCS cluster survey is in still better agreement with the WMAP 3-year
\cite{WMAP3} and COBE 4-year data \cite{COBE4}.  Flux limited
cluster counts can also be used to constrain other cosmological parameters,
such as the equation of state parameter $w$ (for example, \cite{rogerio}).

\subsection{Ineffectiveness of Flux Limited
Cluster Counts for Dimming Mechanisms}
\begin{figure}[h]
\centering
\begin{tabular}{cc}
\epsfig{figure=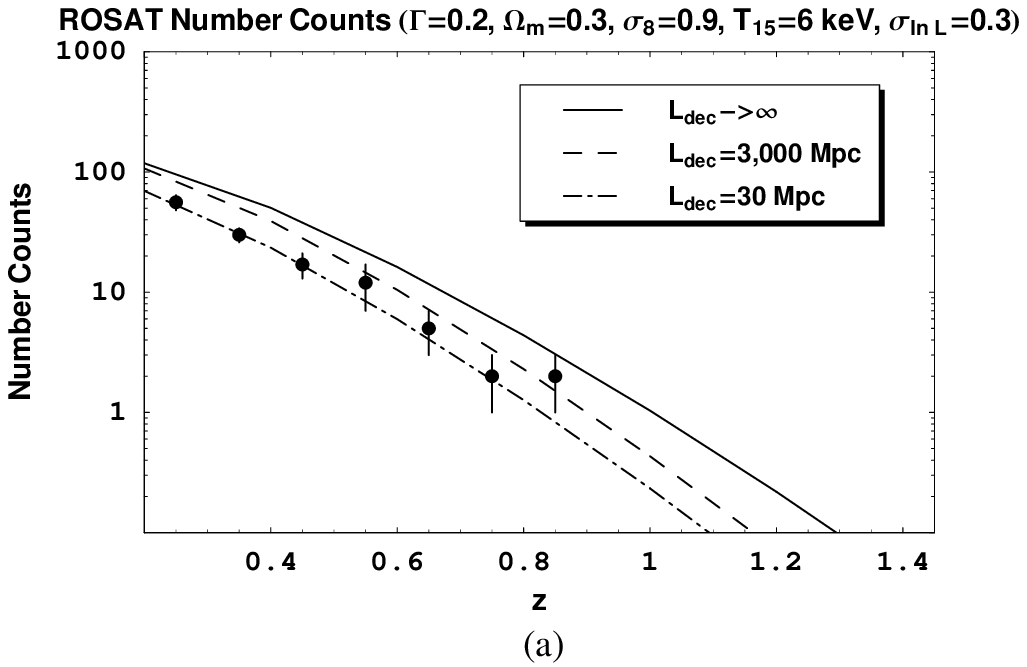,width=3.0in} &
\epsfig{figure=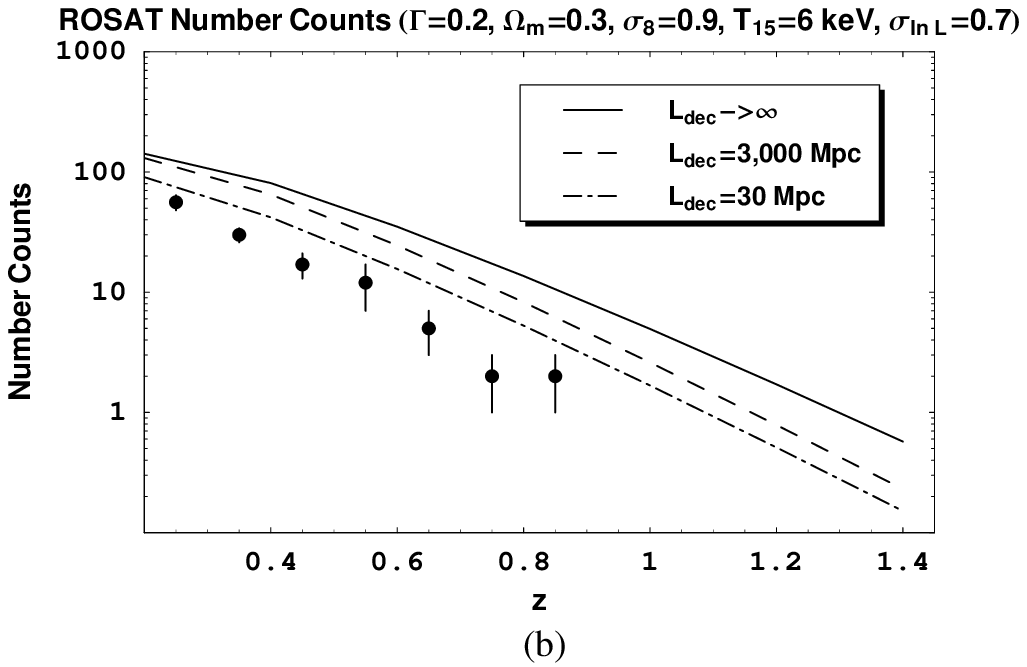,width=3.0in} 
\end{tabular}
\caption{Number Counts
versus redshift for $\Omega_m=0.3$ and $\sigma_8=0.9$ including
axion oscillations for a fixed intrinsic L-T relation, with different
levels of scatter in the L-M
relation. 
(a) $\sigma_{\ln L}=0.3$.  (b) $\sigma_{\ln L}=0.7$.} \label{fig:axplot}
\end{figure}

As we mentioned earlier, we cannot use cluster counts
to constrain dimming mechanisms.  This is not to say that dimming
mechanisms do not affect cluster counts; indeed, dimming would lead to
fewer clusters above the flux limit in any given survey.  
However, the effect of dimming would 
only be through a modification of the observed evolution of cluster
luminosities, which is currently not well constrained theoretically.  
To gauge the effect that
photon-axion oscillations could have on cluster counts we can assume
some particular 
intrinsic $L-T$ evolution, and examine the predicted number counts 
with and without oscillations.  
Figure \ref{fig:axplot} shows number counts versus
redshift for different values of $L_{\rm{dec}}$, where
$L_{\rm{dec}}=\infty$ corresponds to no photon-axion oscillations
and values of $L_{\rm{dec}}$ as low as 30 Mpc correspond to roughly 
$1/3$ of the light lost.  The curves correspond to $\Omega_m=0.3$ and
$\sigma_8=0.9$ with $\Gamma=0.1$ and T$_{15}=6$ keV.  We assumed here
that the intrinsic luminosity evolution is specified by the parameters
(\ref{eq:params}), although there is no theoretical justification for this.

The fact that the observed luminosity evolution is used as input in 
this analysis implies that
the constraints on cosmological parameters from Section~\ref{sec:nonewphysics}
are valid, independent of any dimming mechanism.  As a consequence, 
cluster constraints on the equation of state parameter $w$ can be compared
with constraints from Type Ia supernova surveys, which {\em would be}
affected by dimming mechanisms, as in Refs.~\cite{CKT}.  Such a 
comparison would then provide a new test of the
photon-axion oscillation model.  It would also be interesting to
compare with other constraints on photon-axion oscillations, for example
from CMB spectral distortion \cite{Mirizzi}.

\subsection{Flux Limited Cluster Counts for Other Types of New Physics}
Although we do not attempt further analyses here, flux limited cluster
counts are more suitable for constraining new physics that modifies
structure formation as opposed to the apparent luminosity of clusters.
There are many well-known examples of such possible new physics.
These include possible new interactions in the dark sector and new light
species that would wash out structure. 
Cluster surveys would also be useful in constraining
such phenomena as late time phase transitions in the dark sector and
other aspects of possible new
gravitational dynamics.  It would be straightforward to build these types
of new physics into the formalism described above.

\section{Significance of Future Cluster Surveys}
\label{sec:future}
It is important to recognize that studies of cluster evolution, while already 
interesting, will continue to develop. In this section, we briefly mention some
future approaches that will enhance our knowledge of cluster growth, and note 
the impact of dimming. In particular, Sunyaev-Zeldovich (SZ) surveys such as 
the South Pole Telescope (SPT) \cite{Ruhl:2004kv} or the Atacama Cosmology 
Telescope (ACT) \cite{Kosowsky:2004sw} will establish catalogues of clusters 
which are unbiased in redshift - a crucial element difficult to achieve with 
X-ray surveys. The Dark Energy Survey (DES) \cite{Abbott:2005bi} will take 
advantage of the SPT survey, and include photometric redshifts, as well as 
lensing measurements of cluster masses, and other, independent tests of 
cosmology. The Large Synoptic Survey Telescope (LSST) \cite{Tyson:2003kb} 
will provide masses of a huge set of clusters via weak lensing tomography. 
Future x-ray surveys can expand tremendously the statistics and knowledge of 
many of the uncertainties described 
in earlier sections, in particular the evolution of cluster properties
\cite{Haiman:2005aw}.

One of the key difficulties in using X-ray surveys to extract
cosmology, especially within the present context of photon-axion
oscillations, is the indirect, and uncertain relationship of
luminosity to temperature and temperature to mass. Future studies will
mitigate these issues.

SZ surveys will be a tremendous source of new information in the near future. For a review, see \cite{Birkinshaw:1998qp}. The SZ effect is a decrement $\Delta 
T$ in the CMBR given by the line of sight integral
\begin{equation}
\frac{\Delta T}{T_{CMBR}} = -2 \frac{\sigma_T}{m_e c^2} \int dl \,n_e(l) k_B T_e(l)
\end{equation}
where $\sigma_T$ is the Thomson cross section, $m_e$ is the electron mass, $n_e$ is the electron number density and $T_e$ is the electron temperature. Clusters, with masses in excess of $10^{14} M_\odot$, have sufficient gas densities and temperatures that large scale surveys are possible.
The key feature of the SZ effect is the perturbation of the CMBR which is independent of redshift, and thus allows for a cluster sample, without concern of the selection issues associated with luminosity-weighted X-ray surveys described in earlier sections. 

The SZ effect is not proportional to mass alone, but to the electron pressure. Extracting the mass is a challenge, and of vital importance if these surveys are to provide precision limits on cosmology \cite{Levine:2002uq,Lima:2005tt,Francis:2005nu}. Techniques can involve self calibration  \cite{Majumdar:2003mw,Hu:2003sh,Lima:2005tt,Lima:2007kx}, measurements of the cosmic shear (as in the DES or LSST), or complementary measurements of the cluster x-ray temperature.

As already noted, microwave studies, such as SZ surveys, should not be impacted by dimming mechanisms. Hence, the appearance and properties of clusters within such experiments should be a robust test of the dark energy properties.  Similarly, surveys employing weak lensing will also determine mass and redshift properties will also be insensitive. Supernovae, on the other hand, acting as standard candles, are clearly impacted. As a consequence, studies of cluster growth are key tools of cosmology and measure a quantity {\em distinct} from that of supernovae, quite in the way envisioned by Wang, et al \cite{Wang:2007fsa}.

\section{Conclusions}\label{sec:conclusions}
We have reviewed some of the current models of structure formation and galaxy
cluster dynamics relevant for comparing cluster surveys with models of
particle physics and gravity.  
We compared predictions in the standard $\Lambda$CDM cosmology to the 
400 Square Degree ROSAT galaxy cluster
survey, and found that, with a relatively large assumed scatter in the
relation between cluster mass and temperature, our analysis is consistent
with the WMAP 3-year data.  Earlier analysis of the HIFLUGCS cluster survey
indicates even better agreement with CMB data
\cite{Reiprich}.

We studied a model of 
cluster dimming by photon-axion oscillations, and found that a better
theoretical understanding of cluster luminosity evolution is required 
before firm conclusions could be drawn regarding dimming mechanisms using
cluster data. In particular, improvements in theoretical models and experimental measurements of the evolution of the luminosity-temperature relation may provide an important test of such mechanisms in the future. Moreover, we noted that the cosmological parameters extracted from cluster count surveys are independent of dimming mechanisms, given the measured $L-T$ relation, in contrast to supernovae, and thus provide an independent test of such models. This suggests that such surveys should be folded into analyses such as \cite{Wang:2007fsa} in order to additionally constrain them.
Although cluster counts are insensitive to dimming mechanisms, it is possible to impose relatively strong 
constraints on models of new physics that would affect structure formation
as opposed to cluster appearances. 
This includes any physics that would alter the overall expansion rate of the
universe, the existence of light species that would help to wash out structure,
and additional interactions in the dark sector that could 
either encourage or inhibit the growth of structure.  
The effects of new
physics on structure formation can straightforwardly be built into the 
Press-Schechter formalism.  

The simple
models that enter our fits rely on comparison to numerical simulations for
justification, and deserve to be scrutinized.  
For example, a higher than typical normalization $T_{15}$ in the $M-T$ 
relation, while not justified by simulations,
could bring our fits still better 
in line with the WMAP 3-year data.  Also, the level of
scatter in the $L-T$ and $M-T$ relations deserves to be better analyzed, since
a high assumed scatter brings our fits better in line
with the WMAP fits.  
Despite the remaining uncertainties in this formalism, galaxy cluster surveys
are increasingly ripe for their utilization in constraining new physics.
Upcoming Sunyaev-Zel'dovich and weak-lensing cluster surveys hold the promise of a better
determination of the cluster mass function, and should
help to eliminate some of
the current uncertainties in these techniques.



\section*{Acknowledgments}
J.E. and B.G. thank the NSF
for support under grant PHY-0504442 and the Jeffress Memorial Trust for support under
grant J-768.  NW was supported by NSF CAREER grant PHY-0449818 and DOE grant \# DE-FG02-06ER41417. J.E. thanks the Aspen Center for Physics,
where some of this work was completed.  B.G. thanks the SLAC Summer Institute
for its hospitality while some of this work was completed.


\begin{thebibliography}{999}
\bibitem{WMAP3}
C.~L.~Bennett {\it et al.}  [WMAP Collaboration],
  Astrophys.\ J.\ Suppl.\  {\bf 148}, 1 (2003)
  [arXiv:astro-ph/0302207];
D.~N.~Spergel {\it et al.}  [WMAP Collaboration],
  Astrophys.\ J.\ Suppl.\  {\bf 148}, 175 (2003)
  [arXiv:astro-ph/0302209];
 D.~N.~Spergel {\it et al.}  [WMAP Collaboration],
  Astrophys.\ J.\ Suppl.\  {\bf 170}, 377 (2007)
  [arXiv:astro-ph/0603449].
\bibitem{LSS}
 S.~Perlmutter, M.~S.~Turner and M.~J.~White,
  Phys.\ Rev.\ Lett.\  {\bf 83}, 670 (1999)
  [arXiv:astro-ph/9901052];
 V.~Springel {\it et al.},
  Nature {\bf 435}, 629 (2005)
  [arXiv:astro-ph/0504097].

\bibitem{SNe}
S.~Perlmutter {\it et al.}  [Supernova Cosmology Project Collaboration],
  Astrophys.\ J.\  {\bf 517}, 565 (1999)
  [arXiv:astro-ph/9812133];
A.~G.~Riess {\it et al.}  [Supernova Search Team Collaboration],
  Astrophys.\ J.\  {\bf 607}, 665 (2004)
  [arXiv:astro-ph/0402512].
\bibitem{rot}
 W.~J.~G.~de Blok and S.~S.~McGaugh,
  Mon.\ Not.\ Roy.\ Astron.\ Soc.\  {\bf 290}, 533 (1997)
  [arXiv:astro-ph/9704274];
  S.~S.~McGaugh and W.~J.~G.~de Blok,
  Astrophys.\ J.\  {\bf 499}, 41 (1998)
  [arXiv:astro-ph/9801123].
\bibitem{BBN}
G.~Steigman,
  Int.\ J.\ Mod.\ Phys.\  E {\bf 15}, 1 (2006)
  [arXiv:astro-ph/0511534];
 S.~Burles, K.~M.~Nollett and M.~S.~Turner,
  Phys.\ Rev.\  D {\bf 63}, 063512 (2001)
  [arXiv:astro-ph/0008495].

\bibitem{bullet}
 D.~Clowe, M.~Bradac, A.~H.~Gonzalez, M.~Markevitch, S.~W.~Randall, C.~Jones and D.~Zaritsky,
  arXiv:astro-ph/0608407.

\bibitem{lunar-ranging}
 J.~G.~Williams, X.~X.~Newhall and J.~O.~Dickey,
  Phys.\ Rev.\  D {\bf 53}, 6730 (1996).

\bibitem{rogerio}
 L.~Liberato and R.~Rosenfeld,
  JCAP {\bf 0607}, 009 (2006)
  [arXiv:astro-ph/0604071];
L.~R.~Abramo, R.~C.~Batista, L.~Liberato and R.~Rosenfeld,
  arXiv:0707.2882 [astro-ph].

\bibitem{jain}
B.~Jain and P.~Zhang,
  arXiv:0709.2375 [astro-ph].


\bibitem{PS}
W.~H.~Press and P.~Schechter,
  Astrophys.\ J.\  {\bf 187}, 425 (1974).

\bibitem{cluster-review}
 P.~Rosati, S.~Borgani and C.~Norman,
  Ann.\ Rev.\ Astron.\ Astrophys.\  {\bf 40}, 539 (2002)
  [arXiv:astro-ph/0209035];

\bibitem{Mushotzky}
R.~F.~Mushotzky, Phys. Scr., {\bf T7}, 157 (1984).

\bibitem{David}
 L.~P.~David, A.~Slyz, C.~Jones, W.~Forman, S.~D.~Vrtilek and K.~A.~Arnaud,
  Astrophys.\ J.\  {\bf 412}, 479 (1993).

\bibitem{Markevitch}
M.~Markevitch,
  Astrophys.\ J.\  {\bf 504}, 27 (1998)
  [arXiv:astro-ph/9802059].
\bibitem{Arnaud}
 M.~Arnaud and A.~E.~Evrard,
  Mon.\ Not.\ Roy.\ Astron.\ Soc.\  {\bf 305}, 631 (1999)
  [arXiv:astro-ph/9806353].

\bibitem{Reichart}
D.~E.~Reichart {\it et al.},
  arXiv:astro-ph/9802153.

\bibitem{XMM}
S.~C.~Vauclair {\it et al.},
  Astron.\ Astrophys.\  {\bf 412}, L37 (2003)
  [arXiv:astro-ph/0311381].

\bibitem{Kitzbichler-White}
 M.~G.~Kitzbichler and S.~D.~M.~White,
  Mon.\ Not.\ Roy.\ Astron.\ Soc.\  {\bf 366}, 858 (2006)
  [arXiv:astro-ph/0409682].

\bibitem{Blanchard-Doupsis}
 A.~Blanchard and M.~Douspis,
  arXiv:astro-ph/0405489.

\bibitem{Reiprich}
T.~H.~Reiprich,
  arXiv:astro-ph/0605009.

\bibitem{Wang:2007fsa}
  S.~Wang, L.~Hui, M.~May and Z.~Haiman,
  Phys.\ Rev.\  D {\bf 76}, 063503 (2007)
  [arXiv:0705.0165 [astro-ph]].
  
\bibitem{CKT}
C.~Csaki, N.~Kaloper and J.~Terning,
  Phys.\ Rev.\ Lett.\  {\bf 88}, 161302 (2002)
  [arXiv:hep-ph/0111311];
C.~Csaki, N.~Kaloper and J.~Terning,
  Phys.\ Lett.\  B {\bf 535}, 33 (2002)
  [arXiv:hep-ph/0112212];
J.~Erlich and C.~Grojean,
  Phys.\ Rev.\  D {\bf 65}, 123510 (2002)
  [arXiv:hep-ph/0111335].


\bibitem{SNLS}
 P.~Astier {\it et al.}  [The SNLS Collaboration],
  Astron.\ Astrophys.\  {\bf 447}, 31 (2006)
  [arXiv:astro-ph/0510447].


\bibitem{CKT2}
C.~Csaki, N.~Kaloper and J.~Terning,
  Annals Phys.\  {\bf 317}, 410 (2005)
  [arXiv:astro-ph/0409596].

\bibitem{Maughan}
 B.~J.~Maughan, L.~R.~Jones, H.~Ebeling and C.~Scharf,
  Mon.\ Not.\ Roy.\ Astron.\ Soc.\  {\bf 365}, 509 (2006)
  [arXiv:astro-ph/0503455].


\bibitem{Vikhlinin}
A.~Vikhlinin, L.~Van Speybroeck, M.~Markevitch, W.~R.~Forman and L.~Grego,
  Astrophys.\ J.\  {\bf 578}, L107 (2002)
  [arXiv:astro-ph/0207445].

\bibitem{Mullis}
C.~R.~Mullis {\it et al.},
  Astrophys.\ J.\  {\bf 607}, 175 (2004)
  [arXiv:astro-ph/0401605].

\bibitem{400d}
R.~A.~Burenin, A.~Vikhlinin, A.~Hornstrup, H.~Ebeling, H.~Quintana and A.~Mescheryakov,
  arXiv:astro-ph/0610739.

\bibitem{Review}
T.~Padmanabhan,
in R. Mansouri and R. Brandenberger, {\em Large Scale Structure Formation},
Kluwer Academic Publishers (2000)
  [arXiv:astro-ph/9911374].
Peter Thomas, Lecture Notes,
http://astronomy.sussex.ac.uk/\~petert/galf/notes\_02.pdf

\bibitem{REFLEX}
 H.~Boehringer {\it et al.},
  Astron.\ Astrophys.\  {\bf 425}, 367 (2004)
  [arXiv:astro-ph/0405546].

\bibitem{imperfect}
Tim McKay, Steve Allen, private communication.

\bibitem{K&S}
T.~Kitayama and Y.~Suto,
  Astrophys.\ J.\  {\bf 469}, 480 (1996)
  [arXiv:astro-ph/9604141].

\bibitem{Carroll}
 S.~M.~Carroll, W.~H.~Press and E.~L.~Turner,
  Ann.\ Rev.\ Astron.\ Astrophys.\  {\bf 30}, 499 (1992).

\bibitem{Borgani}
 S.~Borgani, P.~Rosati, P.~Tozzi and C.~Norman,
  arXiv:astro-ph/9901017.

\bibitem{RB}
 T.~H.~Reiprich and H.~Boehringer,
  Astrophys.\ J.\  {\bf 567}, 716 (2002)
  [arXiv:astro-ph/0111285].

\bibitem{dodelson}
S.~Dodelson,
{\em Modern Cosmology},
Amsterdam, Netherlands: Academic Pr. (2003)

\bibitem{EH}
D.~J.~Eisenstein and W.~Hu,
  Astrophys.\ J.\  {\bf 496}, 605 (1998)
  [arXiv:astro-ph/9709112].

\bibitem{Bardeen}
J.~M.~Bardeen, J.~R.~Bond, N.~Kaiser and A.~S.~Szalay,
  Astrophys.\ J.\  {\bf 304}, 15 (1986).

\bibitem{Efstathiou-Bond-White92}
G.~Efstathiou, J.~R.~Bond and S.~D.~M.~White,
  Mon.\ Not.\ Roy.\ Astron.\ Soc.\  {\bf 258}, 1 (1992).

\bibitem{Sugiyama95}
N.~Sugiyama,
  Astrophys.\ J.\ Suppl.\  {\bf 100}, 281 (1995)
  [arXiv:astro-ph/9412025].

\bibitem{SMT}
R.~K.~Sheth, H.~J.~Mo and G.~Tormen,
  Mon.\ Not.\ Roy.\ Astron.\ Soc.\  {\bf 323}, 1 (2001)
  [arXiv:astro-ph/9907024].

\bibitem{Bond}
 J.~R.~Bond, S.~Cole, G.~Efstathiou and N.~Kaiser,
  Astrophys.\ J.\  {\bf 379}, 440 (1991).

\bibitem{Ashfordi}
 N.~Afshordi and R.~Cen,
  Astrophys.\ J.\  {\bf 564}, 669 (2002)
  [arXiv:astro-ph/0105020].

\bibitem{not-isothermal}
 R.~W.~Schmidt and S.~W.~Allen,
  Mon.\ Not.\ Roy.\ Astron.\ Soc.\  {\bf 379}, 209 (2007)
  [arXiv:astro-ph/0610038].


\bibitem{Binney-Tremaine}
J.~Binney and S.~Tremaine, {\em Galactic Dynamics}, Princeton University 
Press, 1987.

\bibitem{NFW}
J.~F.~Navarro, C.~S.~Frenk and S.~D.~M.~White,
  Astrophys.\ J.\  {\bf 490}, 493 (1997)
  [arXiv:astro-ph/9611107];
J.~F.~Navarro, C.~S.~Frenk and S.~D.~M.~White,
  Mon.\ Not.\ Roy.\ Astron.\ Soc.\  {\bf 275}, 720 (1995)
  [arXiv:astro-ph/9408069].

\bibitem{BN}
G.~L.~Bryan and M.~L.~Norman,
  Astrophys.\ J.\  {\bf 495}, 80 (1998)
  [arXiv:astro-ph/9710107].

\bibitem{Oukbir-Blanchard}
J.~Oukbir and A.~Blanchard,
  arXiv:astro-ph/9611085.




\bibitem{Mekal}
 R.~Mewe, E.~H.~B.~Gronenschild and G.~H.~J.~van den Oord,
  Astron.\ Astrophys.\ Suppl.\ Ser.\  {\bf 62}, 197 (1985);
 D.~A.~Liedahl, A.~L.~Osterheld and W.~H.~Goldstein,
  Astrophys.\ J.\  {\bf 438}, L115 (1995).

\bibitem{CIAO}
http://cxc.harvard.edu/ciao/

\bibitem{Jones}
 L.~R.~Jones, C.~Scharf, H.~Ebeling, E.~Perlman, G.~Wegner, M.~Malkan and D.~Horner,
  arXiv:astro-ph/9709189.

\bibitem{Sherpa}
http://cxc.harvard.edu/sherpa/

\bibitem{IGMF}
 A.~Dar and A.~De Rujula,
  Phys.\ Rev.\  D {\bf 72}, 123002 (2005)
  [arXiv:astro-ph/0504480].

\bibitem{Nord}
  R.~Stanek, A.~E.~Evrard, H.~B.~Bohringer, P.~Schuecker and B.~Nord,
  Astrophys.\ J.\  {\bf 648}, 956 (2006)
  [arXiv:astro-ph/0602324].

\bibitem{COBE4}
 E.~F.~Bunn and M.~J.~White,
  Astrophys.\ J.\  {\bf 480}, 6 (1997)
  [arXiv:astro-ph/9607060].

\bibitem{Mirizzi}
 A.~Mirizzi, G.~G.~Raffelt and P.~D.~Serpico,
  Phys.\ Rev.\  D {\bf 72}, 023501 (2005)
  [arXiv:astro-ph/0506078];
A.~Mirizzi, G.~G.~Raffelt and P.~D.~Serpico,
  arXiv:astro-ph/0607415.


\bibitem{Ruhl:2004kv}
  J.~E.~Ruhl {\it et al.}  [The SPT Collaboration],
  arXiv:astro-ph/0411122.
  
\bibitem{Kosowsky:2004sw}
  A.~Kosowsky,
  New Astron.\ Rev.\  {\bf 47}, 939 (2003)
  [arXiv:astro-ph/0402234].

\bibitem{Abbott:2005bi}
  T.~Abbott {\it et al.}  [Dark Energy Survey Collaboration],
  arXiv:astro-ph/0510346.

\bibitem{Tyson:2003kb}
  J.~A.~Tyson  [LSST Collaboration],
  Proc.\ SPIE Int.\ Soc.\ Opt.\ Eng.\  {\bf 4836}, 10 (2002)
  [arXiv:astro-ph/0302102].


\bibitem{Haiman:2005aw}
  Z.~Haiman {\it et al.},
  arXiv:astro-ph/0507013.
  
\bibitem{Birkinshaw:1998qp}
  M.~Birkinshaw,
  Phys.\ Rept.\  {\bf 310}, 97 (1999)
  [arXiv:astro-ph/9808050].
  
\bibitem{Levine:2002uq}
  E.~S.~Levine, A.~E.~Schulz and M.~J.~White,
  Astrophys.\ J.\  {\bf 577}, 569 (2002)
  [arXiv:astro-ph/0204273].
  
\bibitem{Lima:2005tt}
  M.~Lima and W.~Hu,
  Phys.\ Rev.\  D {\bf 72}, 043006 (2005)
  [arXiv:astro-ph/0503363].
  
\bibitem{Francis:2005nu}
  M.~R.~Francis, R.~Bean and A.~Kosowsky,
  JCAP {\bf 0512}, 001 (2005)
  [arXiv:astro-ph/0511161].

\bibitem{Majumdar:2003mw}
  S.~Majumdar and J.~J.~Mohr,
  Astrophys.\ J.\  {\bf 613}, 41 (2004)
  [arXiv:astro-ph/0305341].

\bibitem{Hu:2003sh}
  W.~Hu,
  Phys.\ Rev.\  D {\bf 67}, 081304 (2003)
  [arXiv:astro-ph/0301416].

\bibitem{Lima:2007kx}
  M.~Lima and W.~Hu,
  arXiv:0709.2871 [astro-ph].
  



  
  
  

\end{thebibliography}
\end{document}